\documentclass[preprint,12pt]{elsarticle}
\usepackage[figuresright]{rotating}
\usepackage{amssymb}
\usepackage{amsthm}
\usepackage{multicol}
\usepackage{subfigure}
\usepackage{graphicx}
\usepackage{epstopdf}
\usepackage{fullpage}
\usepackage{latexsym,amsmath}
\usepackage{stmaryrd}
\usepackage{algorithm,algorithmic}
\usepackage{subfigure}
\usepackage{amsfonts}
\usepackage{xcolor,multirow}
\usepackage{bm}
\usepackage{url}
\usepackage{diagbox}
\usepackage{array}
\usepackage{booktabs}
\usepackage{wrapfig}
\biboptions{sort&compress}

\usepackage{amssymb}

\usepackage{tikz}  
\usetikzlibrary{arrows.meta}

\begin{document}

\begin{frontmatter}



\title{Quaternion Matrix Completion Using Untrained Quaternion Convolutional Neural Network for Color Image Inpainting}

\author[lab1]{Jifei Miao}
\ead{jifmiao@163.com}
\author[lab2]{Kit Ian Kou\corref{cor1}}
\ead{kikou@umac.mo}
\author[lab2]{Liqiao Yang}
\ead{liqiaoyoung@163.com}
\author[lab2]{Juan Han}
\ead{juanhan0604@163.com}

\address[lab1]{The School of Mathematics and Statistics, Yunnan University, Kunming, Yunnan, 650091, China}
\address[lab2]{Department of Mathematics, Faculty of
	Science and Technology, University of Macau, Macau 999078, China}
\cortext[cor1]{Corresponding author}

\begin{abstract}
The use of quaternions as a novel tool for color image representation has yielded impressive results in color image processing. By considering the color image as a unified entity rather than separate color space components, quaternions can effectively exploit the strong correlation among the RGB channels, leading to enhanced performance. Especially, color image inpainting tasks are highly beneficial from the application of quaternion matrix completion techniques, in recent years. However, existing quaternion matrix completion methods suffer from two major drawbacks. First, it can be difficult to choose a regularizer that captures the common characteristics of natural images, and sometimes the regularizer that is chosen based on empirical evidence may not be the optimal or efficient option. Second, the optimization process of quaternion matrix completion models is quite challenging because of the non-commutativity of quaternion multiplication. To address the two drawbacks of the existing quaternion matrix completion approaches mentioned above, this paper tends to use an untrained quaternion convolutional neural network (QCNN) to directly generate the completed quaternion matrix. This approach replaces the explicit regularization term in the quaternion matrix completion model with an implicit prior that is learned by the QCNN. Extensive quantitative and qualitative evaluations demonstrate the superiority of the proposed method for color image inpainting compared with some existing quaternion-based and tensor-based methods.
\end{abstract}


\end{frontmatter}

\section{Introduction}
Color image inpainting is used to repair missing or damaged areas of a color image caused by sensor noise, compression artifacts, or other distortion. It can also restore color images with missing regions due to occlusions or other factors. In general, image inpainting can be used to improve the visual quality and completeness of images, and it is a useful tool for many fields including photography, film, and video production, as well as medical imaging and forensics. 

Various methods have been proposed to address color image inpainting, with some of the popular ones including deep learning-based techniques \cite{yu2020region,quan2022image,ran2023multi}, tensor completion methods \cite{qin2022low,he2022tensor,zheng2019low,li2021tensor}, and quaternion matrix completion methods \cite{jia2019robust,miao2021color,chen2022color,jia2022non,miao2020quaternion}. These methods have been extensively researched and have shown significant improvement in the performance of color image inpainting tasks. Although deep learning-based methods often exhibit highly competitive results in color image inpainting, they still have some limitations in some cases. Firstly, training in deep learning methods requires a large amount of labeled data, which may be difficult to obtain in some scenarios. Secondly, deep learning methods often require a significant amount of computational resources and time for training and inference, which may be infeasible for resource-constrained applications that require fast image inpainting. Moreover, since deep learning methods are often trained based on specific data distributions \cite{bengio2013representation}, they may perform poorly under different data distributions. Therefore, non-deep learning methods based on tensor completion and quaternion matrix completion are highly popular in the application of color image inpainting due to their fast computation speed, good interpretability, and excellent performance on small datasets.

\begin{figure}[htbp]
	\centering
	\subfigure[Color pixels in \textit{tensor perspective}.]{
		\includegraphics[width=3.8cm,height=2.5cm]{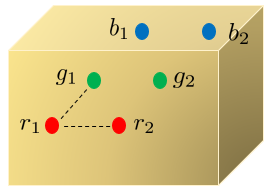}
	}\hspace{2.2cm}
	\subfigure[Color pixels in \textit{quaternion perspective}.]{
		\includegraphics[width=7cm,height=2.7cm]{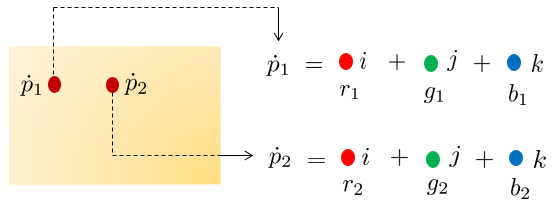}
	}
	\caption{The difference between tensors and quaternions representing color pixels. $r_{m}$, $g_{m}$, and $b_{m}$ respectively denote the RGB channels of the color pixel $\mathbf{p}_{m}=(r_{m},g_{m},b_{m})$ under \textit{tensor perspective} (or $\dot{p}_{m}=r_{m}i+g_{m}j+b_{m}k$ under \textit{quaternion perspective}) for $m=1,2$. The relationship between pixels, \emph{e.g.}, $\mathbf{p}_{1}$ and $\mathbf{p}_{2}$ (or $\dot{p}_{1}$ and $\dot{p}_{2}$ in the quaternion matrix), is called ``spatial relationship'', and the relationship between color channels is called ``intra-channel relationship''. Under \textit{tensor perspective} (a), the intra-channel and spatial relationships are obviously treated equally; that is, the relationship between $r_{1}$ and $g_{1}$ is the same as that between $r_{1}$ and $r_{2}$, which may not be appropriate. Under \textit{quaternion perspective} (b), the three channels are always treated as a whole, and the  intra-channel relationship (bundling with the three imaginary parts of a quaternion) is distinguished from the spatial relationship and can be maintained well.
	}
	\label{TQ_dif}
\end{figure}

Although both third-order tensors and quaternion matrices can be used to represent color images, quaternion matrices, as a novel representation, have more reasonable characteristics and advantages in representing color images. When processing color pixels with RGB channels, third-order tensors may not be able to fully utilize the high correlation among the three channels. This is because the third-order tensors represent color images by simply stacking RGB channels together, which treats the relationship between the RGB channels (referred to as the "intra-channel relationship") and the relationship between pixels (referred to as the "spatial relationship") equally. Figure \ref{TQ_dif}(a) shows the \textit{tensor perspective}. Therefore, any unfolding (matrixization or vectorization) operation of the tensor can break this intra-channel relationship because it is not treated differently from the spatial relationship under the tensor perspective. By contrast, the quaternion always treats the three channels of color pixels as a whole \cite{miao2020low,miao2021color,miao2020quaternion}, so it can preserve this intra-channel relationship well; \emph{see} Figure \ref{TQ_dif}(b), showing the \textit{quaternion perspective}.
Due to the superiority of quaternions in representing color pixels, quaternion matrix completion methods have recently achieved excellent results in color image inpainting. To complete quaternion matrices, there are primarily two approaches: minimizing the nuclear norm of the quaternion matrix \cite{chen2019low,jia2019robust,huang2022quaternion} or decomposing the matrix into low-rank quaternion matrices \cite{miao2021color,miao2020quaternion}. Nevertheless, currently available techniques for completing quaternion matrices have two notable limitations. Firstly, selecting an appropriate regularizer to capture the fundamental features of color images can be problematic, and in some cases, the selected regularizer (\emph{e.g.} rank functions or total variation norm \cite{jia2019color}) based on empirical observations may not be the most efficient or optimal. Secondly, the optimization process for quaternion matrix completion models is arduous due to the non-commutative nature of quaternion multiplication. Consequently, this paper aims to overcome the limitations of the current quaternion matrix completion methods mentioned above by utilizing an untrained quaternion convolutional neural network (QCNN) to produce the completed quaternion matrix directly. The proposed method has the following advantages: $1)$ Compared with traditional deep learning-based methods, our method directly exploits the deep priors of color images \cite{ulyanov2018deep}, and does not require a large amount of data to pre-train the network. $2)$ Using a network based on QCNN \cite{DBLP:journals/air/ParcolletML20,DBLP:conf/ijcnn/GaudetM18,parcollet2018quaternion} to generate color images has advantages over traditional CNN, including prevention of overfitting, fewer parameters, and most importantly, the ability to fully simulate the inherent relationships between color image channels. A detailed explanation of QCNN can be found in Section \ref{qcnndt}. $3)$ This method will effectively avoid the limitations of existing quaternion matrix completion methods. Specifically, the QCNN learns an implicit prior that replaces the explicit regularization term in the quaternion matrix completion model, which  will liberate researchers from the anguish of searching for suitable regularization terms and designing complex optimization algorithms for quaternion matrix completion models.

The remaining chapters of this paper are organized as follows: Section \ref{sec:2} introduces quaternions and QCNN. Section \ref{sec:3} provides details of the proposed color image inpainting approach. The qualitative and quantitative experiments are presented and analyzed in Section \ref{sec:4}. Finally, some conclusions are drawn in Section \ref{sec:5}. 

\section{Quaternions and Quaternion Convolutional Neural Network}
\label{sec:2}
In this section, we will provide a concise overview of quaternion algebra before delving into a thorough analysis of the key components that make up the QCNN.
\subsection{Quaternion Algebra}
As a natural extension of the complex domain, a quaternion $\dot{q}\in\mathbb{H}$ consisting of one real part and three imaginary parts is defined as
\begin{equation}
	\label{equ1}
	\dot{q}=\underbrace{q_{0}}_{{\rm{Re}}(\dot{q})}+\underbrace{q_{1}\texttt{i}+q_{2}\texttt{j}+q_{3}\texttt{k}}_{{\rm{Im}}(\dot{q})},
\end{equation}
where $q_{l}\in\mathbb{R}\: (l=0,1,2,3)$, and $\texttt{i}, \texttt{j}, \texttt{k}$ are
imaginary number units and obey the quaternion rules that
\begin{equation*}
	\left\{
	\begin{array}{lc}
		\texttt{i}^{2}=\texttt{j}^{2}=\texttt{k}^{2}=\texttt{i}\texttt{j}\texttt{k}=-1,\\
		\texttt{i}\texttt{j}=-\texttt{j}\texttt{i}=\texttt{k}, \texttt{j}\texttt{k}=-\texttt{k}\texttt{j}=\texttt{i}, \texttt{k}\texttt{i}=-\texttt{i}\texttt{k}=\texttt{j}.
	\end{array}
	\right.
\end{equation*}
If the real part $q_{0}:={\rm{Re}}(\dot{q})=0$, then  $\dot{q}=q_{1}\texttt{i}+q_{2}\texttt{j}+q_{3}\texttt{k}:={\rm{Im}}(\dot{q})$ is named a pure quaternion.
The conjugate and the modulus of a quaternion $\dot{q}$ are,
respectively, defined as 
\begin{equation*}
	\dot{q}^{\ast}=q_{0}-q_{1}\texttt{i}-q_{2}\texttt{j}-q_{3}\texttt{k} \quad  \text{and} \quad 
	|\dot{q}|=\sqrt{q_{0}^{2}+q_{1}^{2}+q_{2}^{2}+q_{3}^{2}}.
\end{equation*}
Given two quaternions $\dot{p}$ and $\dot{q}\in\mathbb{H}$, the multiplication of them is
\begin{equation}
\label{equ2}
\begin{split}
	\dot{p}\dot{q}=&(p_{0}q_{0}-p_{1}q_{1}-p_{2}q_{2}-p_{3}q_{3})\\
	&+(p_{0}q_{1}+p_{1}q_{0}+p_{2}q_{3}-p_{3}q_{2})\texttt{i}\\
	&+(p_{0}q_{2}-p_{1}q_{3}+p_{2}q_{0}+p_{3}q_{1})\texttt{j}\\
	&+(p_{0}q_{3}+p_{1}q_{2}-p_{2}q_{1}+p_{3}q_{0})\texttt{k},
\end{split}
\end{equation}
which is also referred to as Hamilton product \cite{DBLP:journals/air/ParcolletML20}. Analogously, a quaternion matrix $\dot{\mathbf{Q}}=(\dot{q}_{mn})\in\mathbb{H}^{M\times N}$ is written
as $\dot{\mathbf{Q}}=\mathbf{Q}_{0}+\mathbf{Q}_{1}\texttt{i}+\mathbf{Q}_{2}\texttt{j}+\mathbf{Q}_{3}\texttt{k}$, where $\mathbf{Q}_{l}\in\mathbb{R}^{M\times N}\: (l=0,1,2,3)$, $\dot{\mathbf{Q}}$ is named a pure quaternion matrix when $\mathbf{Q}_{0}={\rm{Re}}(\dot{\mathbf{Q}})=\mathbf{0}$.

\subsection{Quaternion Convolutional Neural Networks}
\label{qcnndt}
QCNN has quaternionic model parameters, inputs, activations, and outputs. In the following, we recall and analyze the key components used in this paper of QCNN, \emph{e.g.}, quaternion convolution, quaternion activation functions, and quaternion-valued backpropagation.
\subsubsection{Quaternion Convolution}
\begin{figure}[htbp]
	\begin{minipage}{0.4\linewidth}
		\centerline{\includegraphics[width=4cm]{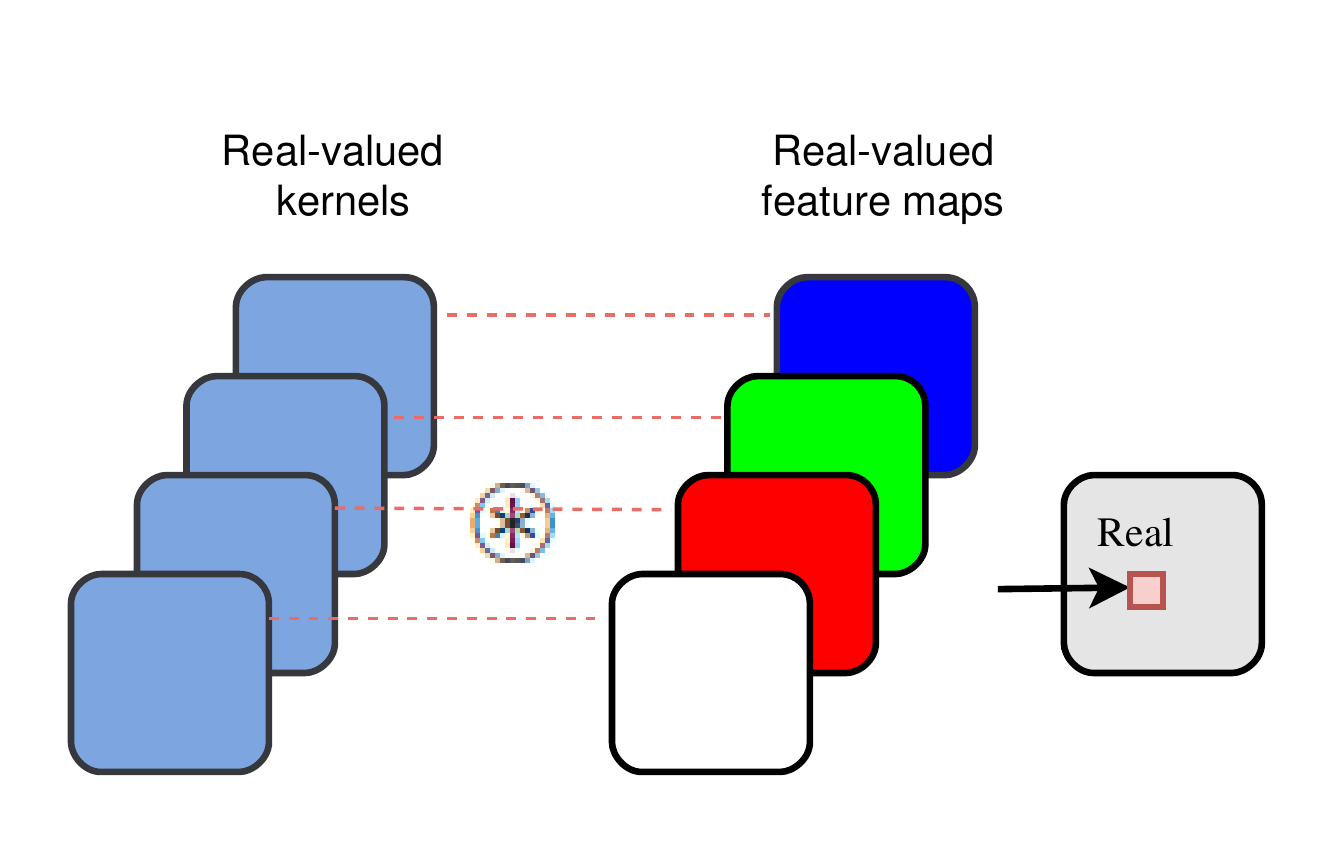}}
		\centerline{(a) Real-valued convolution}
	\end{minipage}
	\begin{minipage}{0.5\linewidth}
		\centerline{\includegraphics[width=10.2cm]{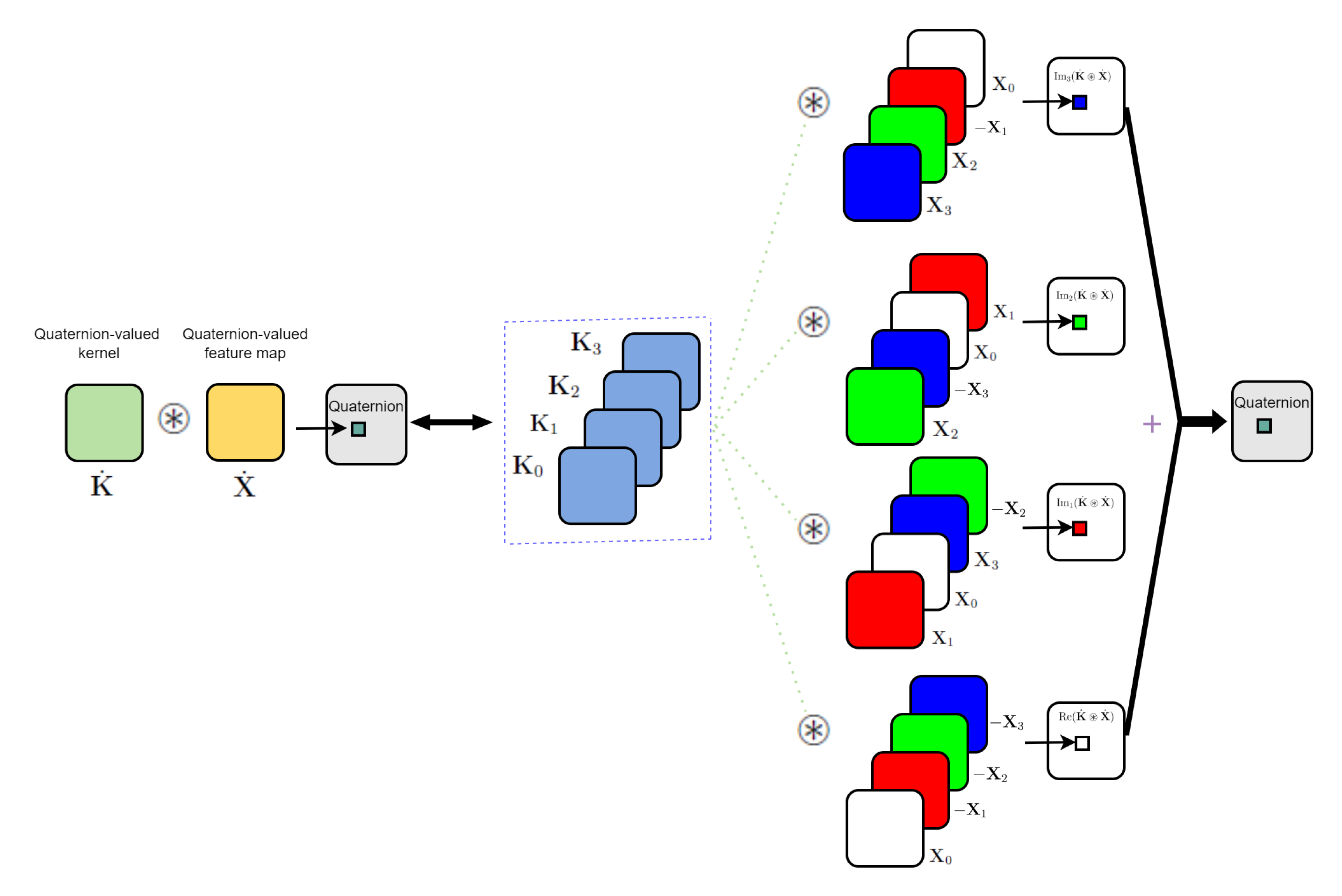}}
		\centerline{(b) Quaternion convolution}
	\end{minipage}
\caption{The differences between (a) real-valued convolution and (b) quaternion convolution.}
\label{fig1}
\end{figure}

Convolution in the quaternion domain formally can be defined the same as that in the real domain \cite{DBLP:conf/ijcnn/GaudetM18,DBLP:conf/interspeech/ParcolletZMTLMB18,9274493Zhou}. Letting $\dot{\mathbf{K}}=(\dot{k}_{mn})=\mathbf{K}_{0}+\mathbf{K}_{1}\texttt{i}+\mathbf{K}_{2}\texttt{j}+\mathbf{K}_{3}\texttt{k}$  be a quaternion convolution kernel matrix, and $\dot{\mathbf{Y}}=(\dot{y}_{mn})=\mathbf{Y}_{0}+\mathbf{Y}_{1}\texttt{i}+\mathbf{Y}_{2}\texttt{j}+\mathbf{Y}_{3}\texttt{k}$  be a quaternion input matrix, their convolution in deep learning is computed as 
\begin{equation}\label{equ3}
(\dot{\mathbf{K}}\circledast\dot{\mathbf{Y}})(r_{1},r_{2})=\sum_{m}\sum_{n}\dot{k}_{mn}\dot{y}_{r_{1}+m,r_{2}+n},
\end{equation}
where $\circledast$ denotes convolution operation. Deconvolution, strided convolution, dilated convolution, and padding in quaternion domain are also defined analogously to real-valued convolution. Assume that $\dot{\mathbf{X}}=\mathbf{X}_{0}+\mathbf{X}_{1}\texttt{i}+\mathbf{X}_{2}\texttt{j}+\mathbf{X}_{3}\texttt{k}$ is a certain window (patch) of $\dot{\mathbf{Y}}$, and has the same size as $\dot{\mathbf{K}}$. Based on the Hamilton product (\ref{equ2}), the convolution of $\dot{\mathbf{K}}$ and $\dot{\mathbf{X}}$ can be written as
\begin{equation}
	\label{equ4}
	\begin{split}
		\dot{\mathbf{K}}\circledast\dot{\mathbf{X}}=&(\mathbf{K}_{0}\circledast\mathbf{X}_{0}-\mathbf{K}_{1}\circledast\mathbf{X}_{1}-\mathbf{K}_{2}\circledast\mathbf{X}_{2}-\mathbf{K}_{3}\circledast\mathbf{X}_{3})\\
		&+(\mathbf{K}_{0}\circledast\mathbf{X}_{1}+\mathbf{K}_{1}\circledast\mathbf{X}_{0}+\mathbf{K}_{2}\circledast\mathbf{X}_{3}-\mathbf{K}_{3}\circledast\mathbf{X}_{2})\texttt{i}\\
		&+(\mathbf{K}_{0}\circledast\mathbf{X}_{2}-\mathbf{K}_{1}\circledast\mathbf{X}_{3}+\mathbf{K}_{2}\circledast\mathbf{X}_{0}+\mathbf{K}_{3}\circledast\mathbf{X}_{1})\texttt{j}\\
		&+(\mathbf{K}_{0}\circledast\mathbf{X}_{3}+\mathbf{K}_{1}\circledast\mathbf{X}_{2}-\mathbf{K}_{2}\circledast\mathbf{X}_{1}+\mathbf{K}_{3}\circledast\mathbf{X}_{0})\texttt{k}.
	\end{split}
\end{equation} 
From (\ref{equ4}), one can see that if $\dot{\mathbf{X}}$ and $\dot{\mathbf{K}}$ are real-valued matrices, \emph{i.e.}, $\mathbf{X}_{1}=\mathbf{X}_{2}=\mathbf{X}_{3}=\mathbf{K}_{1}=\mathbf{K}_{2}=\mathbf{K}_{3}=\mathbf{0}$, then the convolution degrades into the traditional real-valued case. Labeling $\dot{\mathbf{K}}\circledast\dot{\mathbf{X}}={\rm{Re}}(\dot{\mathbf{K}}\circledast\dot{\mathbf{X}})+{\rm{Im}}_{1}(\dot{\mathbf{K}}\circledast\dot{\mathbf{X}})\texttt{i}+{\rm{Im}}_{2}(\dot{\mathbf{K}}\circledast\dot{\mathbf{X}})\texttt{j}+{\rm{Im}}_{3}(\dot{\mathbf{K}}\circledast\dot{\mathbf{X}})\texttt{k}$, and using a matrix to represent the components of the convolution, we express the quaternion convolution formula (\ref{equ4}) in the following way similar to matrix multiplication:

\begin{equation}\label{equ5}
\begin{bmatrix} 
	{\rm{Re}}(\dot{\mathbf{K}}\circledast\dot{\mathbf{X}}) \\ {\rm{Im}}_{1}(\dot{\mathbf{K}}\circledast\dot{\mathbf{X}})\\{\rm{Im}}_{2}(\dot{\mathbf{K}}\circledast\dot{\mathbf{X}})\\{\rm{Im}}_{3}(\dot{\mathbf{K}}\circledast\dot{\mathbf{X}})
\end{bmatrix}=
\begin{bmatrix}
\mathbf{K}_{0}&	-\mathbf{K}_{1}&-\mathbf{K}_{2}&-\mathbf{K}_{3}\\
\mathbf{K}_{1}&	\mathbf{K}_{0}&-\mathbf{K}_{3}&\mathbf{K}_{2}\\
\mathbf{K}_{2}&	\mathbf{K}_{3}&\mathbf{K}_{0}&-\mathbf{K}_{1}\\
\mathbf{K}_{3}&	-\mathbf{K}_{2}&\mathbf{K}_{1}&\mathbf{K}_{0}
\end{bmatrix}\circledast
\begin{bmatrix} 
\mathbf{X}_{0}\\ \mathbf{X}_{1}\\ \mathbf{X}_{2}\\ \mathbf{X}_{3}
\end{bmatrix}. 
\end{equation}
In addition, one can visually see the differences between real-valued convolution and quaternion convolution in Figure \ref{fig1}. From formula (\ref{equ5}) and Figure \ref{fig1}, we can notice that the quaternion convolution forces each component of the quaternion kernel $\dot{\mathbf{K}}$ to interact
with each component of the input quaternion feature map $\dot{\mathbf{X}}$. This kind of interaction mechanism forces the kernel to capture internal latent relations among different channels of the feature map since each characteristic in a channel will have an influence on the other channels through the common kernel. Different from real-valued convolution, which simply multiplies each kernel with the corresponding feature map, the quaternion convolution is similar to a mixture of standard convolution. Such a mixture can perfectly simulate the potential relationship between color image channels. This may be the fundamental reason why QCNN is more suitable for generating color images.
No real-valued CNN would make such connections without the inspiration from quaternion convolutions, although it is feasible to incorporate this mixture into three real-valued CNNs using supplementary connections. Furthermore, when the quaternion convolution layer has the same output dimensions (a quaternion has four dimensions) as the real-valued convolution layer, for the quaternion convolution, the parameters that need to be learned are only  $\frac{1}{4}$ of the real-valued convolution, which has great potential to avoid the over-fitting phenomenon. These exciting characteristics of quaternion convolution are our main motivation for designing QCNN to color image inpainting tasks.


\subsubsection{Quaternion activation functions}
Many quaternion activation functions have been investigated, whereas the split activation \cite{DBLP:journals/air/ParcolletML20,DBLP:conf/asru/ParcolletML17}, a more frequent and simpler solution, is applied
in our proposed model. Let $Q_\mathfrak{f}(\dot{\mathbf{Y}})$ be a split activation function applied to the quaternion $\dot{\mathbf{Y}}=\mathbf{Y}_{0}+\mathbf{Y}_{1}\texttt{i}+\mathbf{Y}_{2}\texttt{j}+\mathbf{Y}_{3}\texttt{k}$, such that
\begin{equation*}
Q_\mathfrak{f}(\dot{\mathbf{Y}})=\mathfrak{f}(\mathbf{Y}_{0})+\mathfrak{f}(\mathbf{Y}_{1})\texttt{i}+\mathfrak{f}(\mathbf{Y}_{2})\texttt{j}+\mathfrak{f}(\mathbf{Y}_{3})\texttt{k},	
\end{equation*}
with $\mathfrak{f}$ corresponding to any traditional real-valued activation function, \emph{e.g.}, ReLU, LeakyReLU, sigmoid, \emph{etc.} Thus, $Q_\mathfrak{f}$ can be, $Q_{ReLU}$, $Q_{LeakyReLU}$, $Q_{sigmoid}$, $Q_{Tanh}$, \emph{etc.} 
\subsubsection{Quaternion-Valued Backpropagation}
The quaternion-valued backpropagation is just an extension of the method for its real-valued counterpart \cite{DBLP:journals/air/ParcolletML20}. The gradient of a general quaternion loss function $\mathcal{L}$ is computed for each component of the quaternion kernel matrix $\dot{\mathbf{K}}$ as 
\begin{equation*}
\frac{\nabla\mathcal{L}}{\nabla\dot{\mathbf{K}}}=\frac{\nabla\mathcal{L}}{\nabla\mathbf{K}_{0}}+\frac{\nabla\mathcal{L}}{\nabla\mathbf{K}_{1}}\texttt{i}+\frac{\nabla\mathcal{L}}{\nabla\mathbf{K}_{2}}\texttt{j}+\frac{\nabla\mathcal{L}}{\nabla\mathbf{K}_{3}}\texttt{k},
\end{equation*}
where $\nabla$ denotes gradient operator.
Afterwards, the gradient is propagated back based on the chain rule. Thus, QCNN can be easily trained
as real-valued CNN following the backpropagation.

\section{Color Image Inpainting}
\label{sec:3}
\begin{figure}[htbp]
	\centering
	\includegraphics[width=13.5cm,height=3cm]{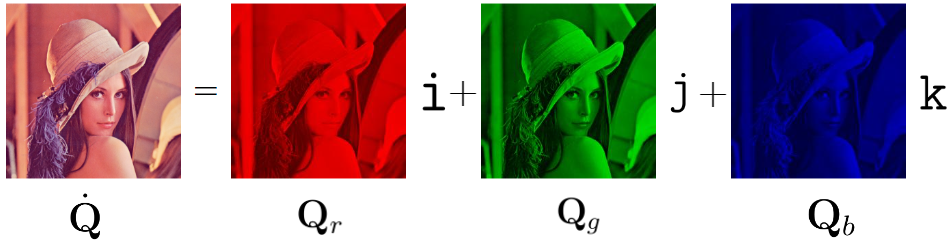}
	\caption{Color image represented by quaternion matrix.}
	\label{rgb}
\end{figure}

As quaternions offer a superior method of representing color pixels, every pixel in an RGB color image can be encoded as a pure quaternion. That is
\begin{equation}
	\label{equ6_1}
	\dot{q}=0+q_{r}\texttt{i}+q_{g}\texttt{j}+q_{b}\texttt{k},
\end{equation}
where $\dot{q}$ denotes a color pixel, $q_{r}$, $q_{g}$, and $q_{b}$ are respectively the pixel values in red, green, and blue channels. Naturally, the given color image with spatial resolution of $M\times N$ pixels can be represented by a pure quaternion matrix $\dot{\mathbf{Q}}=(\dot{q}_{mn})\in\mathbb{H}^{M\times N}$, $1\leq m\leq M$, $1\leq n\leq N$ as follows:
\begin{equation}
	\label{quaternion_rc}
	\dot{\mathbf{Q}}=\mathbf{0}+\mathbf{Q}_{r}\texttt{i}+\mathbf{Q}_{g}\texttt{j}+\mathbf{Q}_{b}\texttt{k},
\end{equation}
where $\mathbf{Q}_{r}, \mathbf{Q}_{g}, \mathbf{Q}_{b}\in\mathbb{R}^{M\times N}$ containing respectively  the pixel values in red, green, and blue channels. Figure \ref{rgb} shows an example of using a quaternion matrix to represent a color image.

\subsection{Optimization Model}
In the quaternion domain, a general quaternion matrix completion model for color image inpainting can be developed as 
\begin{equation}
	\label{mode1}
		\mathop{{\rm{min}}}\limits_{\dot{\mathbf{X}}}\ \frac{1}{2} \|\mathcal{P}_{\Omega}(\dot{\mathbf{X}}-\dot{\mathbf{Q}})\|_{F}^{2}+\lambda\Phi(\dot{\mathbf{X}}),
\end{equation}
where $\lambda$ is a nonnegative parameter, $\dot{\mathbf{X}}\in\mathbb{H}^{M\times N}$ is a desired output completed quaternion matrix, $\dot{\mathbf{Q}}\in\mathbb{H}^{M\times N}$ is an observed quaternion matrix with missing entries, $\Phi(\cdot)$ is a regularization operator, and $\mathcal{P}_{\Omega}$ is the unitary projection onto the linear space of matrices supported on the entries set $\Omega$, defined as
\begin{equation*}
	(\mathcal{P}_{\Omega}(\dot{\mathbf{X}}))_{mn}=\left\{
	\begin{array}{c}
		\!\!\!\dot{x}_{mn},\qquad (m,n)\in \Omega, \\
		0,\qquad\quad\:  (m,n)\notin\Omega.
	\end{array}
	\right.
\end{equation*}
In model (\ref{mode1}), $\Phi(\cdot)$ can be the rank function, as used in recent quaternion matrix completion models \cite{chen2019low,song2021low,yang2021weighted}, or any other suitable regularizer, or a combination of them. However, selecting an appropriate regularizer to capture the generic prior of natural images can be challenging, and in some cases, the empirically chosen regularizer may not be the most suitable or effective one. Furthermore, optimizing regularizers in the quaternion domain is a challenging task because of the non-commutativity of quaternion multiplication. Thus, in this paper, we propose to replace the explicit regularization term $\Phi(\cdot)$ with an implicit prior learned by the QCNN. As a result, the model (\ref{mode1}) becomes
\begin{equation}
	\label{mode2}
	\mathop{{\rm{min}}}\limits_{\dot{\mathbf{\theta}}}\  \|\mathcal{P}_{\Omega}(f_{\dot{\mathbf{\theta}}}(\dot{\mathbf{Z}})-\dot{\mathbf{Q}})\|_{F}^{2},\quad  \text{and} \quad  \dot{\mathbf{X}}_{opt}=f_{\dot{\mathbf{\theta}}_{opt}}(\dot{\mathbf{Z}}),
\end{equation}
where $\dot{\mathbf{Z}}$, a random initialization with the same size as $\dot{\mathbf{Q}}$, is passed as input to the QCNN $f_{\dot{\mathbf{\theta}}}(\dot{\mathbf{Z}})$ which is parameterized by $\dot{\mathbf{\theta}}$. Since there should not be any change in the uncorrupted regions, the final output pixels outside of the corrupted areas are replaced with the original input values. Therefore,
once we get $\dot{\mathbf{X}}_{opt}$, the final inpainted color image is
\begin{equation}\label{mode3}
\dot{\mathbf{X}}_{inpainted}=\mathcal{P}_{\Omega}(\dot{\mathbf{Q}})+\mathcal{P}_{\Omega^{c}}(\dot{\mathbf{X}}_{opt}),	
\end{equation}
where $\Omega^{c}$  is the complement of $\Omega$.

\subsection{The Designed QCNN}
\begin{figure}[htbp]
	\centering
	\includegraphics[width=13cm,height=5cm]{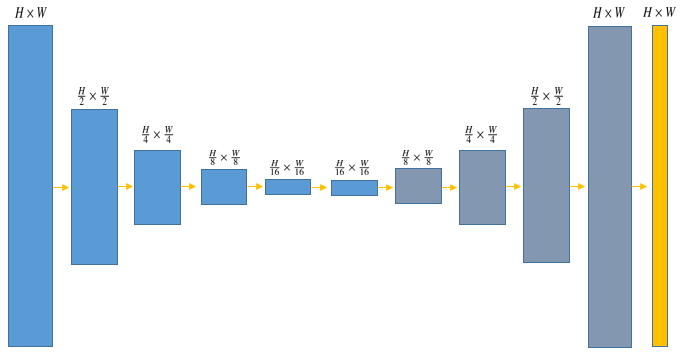}
	\caption{The architecture of the designed QCNN.}
	\label{qcnnarc}
\end{figure}

The designed QCNN is an encoder-decoder, which includes an encoding part with four times downsampling and a decoding part with four quaternion deconvolution (Qdeconv) layers for backing to the original size of the color image. The architecture and details of the designed QCNN can be seen in Figure \ref{qcnnarc} and Table \ref{tab1}, respectively. The quaternion batch normalization (QBN) method studied in \cite{DBLP:conf/ijcnn/GaudetM18,DBLP:journals/access/YinWLZJS19} is also used in the designed QCNN to stabilize and speed up the process of generating the quaternion matrix $\dot{\mathbf{X}}_{opt}$.
\begin{table}[htbp]
\centering
\caption{The details of the designed QCNN.}   
\begin{tabular}{lcccl}    
\toprule Module types&Kernel size &Stride &Output channels \\   
\midrule
Qconv+QBN+$Q_{LeakyReLU}$&$3\times 3$&$1\times 1$ & $64$ \\     
Qconv+QBN+$Q_{LeakyReLU}$&$3\times 3$&$2\times 2$ & $64$ \\   
Qconv+QBN+$Q_{LeakyReLU}$&$3\times 3$&$2\times 2$ & $64$ \\   
Qconv+QBN+$Q_{LeakyReLU}$&$3\times 3$&$2\times 2$ & $64$ \\   
Qconv+QBN+$Q_{LeakyReLU}$&$3\times 3$&$2\times 2$ & $64$ \\ 
  
Qconv+QBN+$Q_{LeakyReLU}$&$3\times 3$&$1\times 1$ & $64$ \\ 
    
Qdeconv+QBN+$Q_{LeakyReLU}$&$3\times 3$&$2\times 2$ & $64$ \\   
Qdeconv+QBN+$Q_{LeakyReLU}$&$3\times 3$&$2\times 2$ & $64$ \\   
Qdeconv+QBN+$Q_{LeakyReLU}$&$3\times 3$&$2\times 2$ & $64$ \\   
Qdeconv+QBN+$Q_{LeakyReLU}$&$3\times 3$&$2\times 2$ & $64$ \\   
Qconv&$3\times 3$&$1\times 1$ & $1$ \\   
\bottomrule    
\end{tabular}
\label{tab1}  
\end{table}

\section{Experiments}
\label{sec:4}
\subsection{Experiment Settings}
The experiments for our proposed QCNN-based method for color image inpainting were conducted in an environment with ``torch==1.13.1+cu116''. During backpropagation, we have chosen the Adam optimizer with a learning rate of 0.01. In addition, the input of the QCNN, $\dot{\mathbf{Z}}$, is a random quaternion matrix with the same spatial size as the color image to be inpainted. The output of the QCNN is a completed quaternion matrix whose imaginary parts correspond to the inpainted RGB color image. 

\subsection{Datasets}
\begin{figure*}[htbp]
	\centering
	\subfigure[baboon]{
		\includegraphics[width=3.2cm,height=2cm]{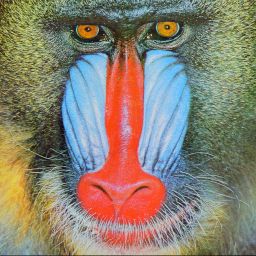}
	}
	\subfigure[monarch]{
		\includegraphics[width=3.2cm,height=2cm]{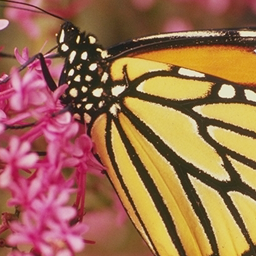}
	}
	\subfigure[airplane]{
		\includegraphics[width=3.2cm,height=2cm]{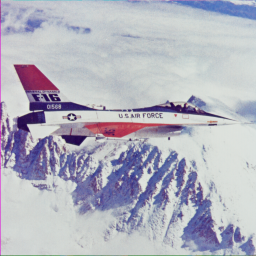}
	}
	\subfigure[peppers]{
		\includegraphics[width=3.2cm,height=2cm]{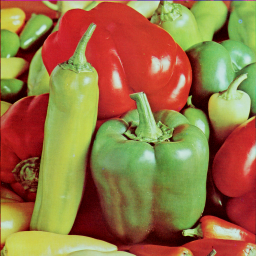}
	}\\
	\subfigure[sailboat]{
		\includegraphics[width=3.2cm,height=2cm]{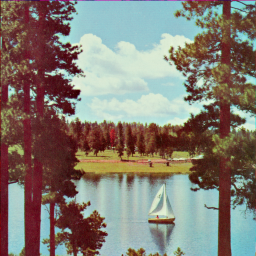}
	}
	\subfigure[lena]{
		\includegraphics[width=3.2cm,height=2cm]{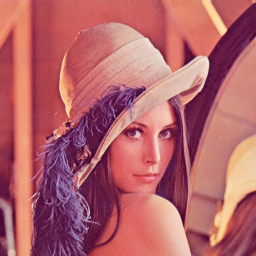}
	}
	\subfigure[panda]{
		\includegraphics[width=3.2cm,height=2cm]{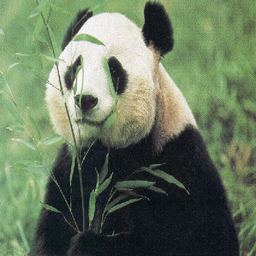}
	}
	\subfigure[barbara]{
		\includegraphics[width=3.2cm,height=2cm]{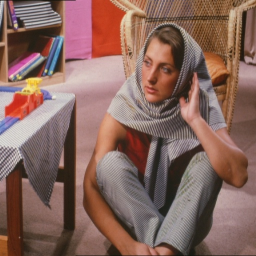}
	}
	\caption{Tested color images.}
	\label{Test_data}
\end{figure*}

We evaluate the proposed method on eight common-used color images (including ``baboon'', ``monarch'', ``airplane'', ``peppers'', ``sailboat'', ``lena'', ``panda'', and ``barbara'', \emph{see} Figure \ref{Test_data}) with spatial size $256\times 256$.
For random missing, we set three levels of sampling rates (SRs) which are ${\rm{SR}}=10\%$, ${\rm{SR}}=30\%$, and ${\rm{SR}}=50\%$.
For structural missing, we use two kinds of cases, \emph{see} the observed color images in Figure \ref{im_Es5}.

\subsection{Comparison}
We compare our QCNN-based method with its real-domain counterpart, \emph{i.e.}, the CNN-based method with the same number of learnable parameters as our model. Additionally, we compare our method with several well-known tensor and quaternion-based techniques that utilize low-rank regularization, namely t-SVD \cite{DBLP:journals/tsp/ZhangA17}, TMac-TT \cite{DBLP:journals/tip/BenguaPTD17}, LRQA-2 \cite{chen2019low}, LRQMC \cite{miao2021color}, and TQLNA \cite{DBLP:journals/isci/YangMK22}. The implementation of all the comparison methods follow their original papers and source codes. 
For measuring the quality of the results, two metrics, peak signal-to-noise ratio (PSNR) and  structure similarity index (SSIM) \cite{DBLP:journals/tip/WangBSS04} are used in this paper.
\begin{table}[htbp]\tiny
	\centering
	\caption{The PSNR and SSIM values of different methods on the eight color images with three levels of sampling rates (the format is PSNR/SSIM, and \textbf{bold} fonts denote the best performance).}   
\centering
\resizebox{16.4cm}{6cm}{
	\begin{tabular}{|c|c|c|c|c|c|c|c|}		
		\hline
		Methods:&CNN &t-SVD \cite{DBLP:journals/tsp/ZhangA17} & TMac-TT \cite{DBLP:journals/tip/BenguaPTD17}& LRQA-2 \cite{chen2019low}& LRQMC \cite{miao2021color} &TQLNA \cite{DBLP:journals/isci/YangMK22} &\textbf{Ours} \\ \toprule
		\hline
		Images:  &\multicolumn{7}{c|}{${\rm{SR}}=10\%$}\\
		\hline
		baboon &18.456/0.572&17.325/0.505&19.242/0.579&17.941/0.527&18.065/0.546&18.075/0.537&\textbf{20.301}/\textbf{0.647}\\
		monarch&19.589/0.860&17.325/0.643&17.306/0.763&14.561/0.642&14.752/0.684&13.534/0.579 &\textbf{20.292}/\textbf{0.874}  \\
		airplane&21.001/0.675&17.765/0.377&20.503/0.588&18.601/0.424&18.715/0.480&18.442/0.416   &\textbf{22.258}/\textbf{0.715}  \\
		peppers&23.328/0.929&15.993/0.718&20.416/0.871&17.111/0.761&16.206/0.740&17.145/0.764  &\textbf{23.672}/\textbf{0.932} \\  
		sailboat&20.018/0.781&16.514/0.502&17.606/0.670&17.103/0.550&17.295/0.581&16.711/0.526&\textbf{20.714}/\textbf{0.785}  \\
		lena&22.773/0.913&17.659/0.798&21.616/0.877&18.607/0.811&18.553/0.825&18.528/0.802&\textbf{23.724}/\textbf{0.925} \\
		panda&23.399/0.695&18.074/0.492&22.518/0.628&19.205/0.514&19.212/0.562&19.149/0.513     &\textbf{24.408}/\textbf{0.764}\\
		barbara&21.524/0.754&16.894/0.513&20.373/0.723&17.917/0.533&17.947/0.585&17.996/0.527  &\textbf{22.943}/\textbf{0.783}\\
		\hline
		\hline
		Images:  &\multicolumn{7}{c|}{${\rm{SR}}=30\%$}\\
		\hline
		baboon&21.800/0.761 &20.657/0.703&21.801/0.751&20.685/0.695&21.279/0.727&20.878/0.705&\textbf{22.565}/\textbf{0.775}\\
		monarch&25.163/0.950 &19.003/0.833&22.487/0.910&19.582/0.842&19.725/0.851&19.876/0.848&\textbf{25.278}/\textbf{0.952}\\
		airplane &25.301/0.843&22.555/0.671&24.096/0.821&22.982/0.681&23.183/0.724&23.250/0.702&\textbf{26.119}/\textbf{0.852}\\
		peppers&28.281/0.973 &22.287/0.908&25.458/0.951&23.330/0.923&23.671/0.930&23.976/0.934&\textbf{28.536}/\textbf{0.976}\\  
		sailboat&24.374/0.904&20.958/0.767&22.819/0.862&21.343/0.779&21.634/0.806&21.609/0.792&\textbf{24.723}/\textbf{0.906}\\
		lena&27.519/0.963&23.217/0.917&26.136/0.951&23.729/0.921&24.173/0.931&24.059/0.927&\textbf{27.730}/\textbf{0.966} \\
		panda&27.416/0.850 &23.698/0.730&26.602/0.837&24.297/0.739&24.479/0.772&24.721/0.753&\textbf{27.828}/\textbf{0.862}\\
		barbara&25.804/0.875 &22.737/0.771&25.338/0.866&23.403/0.781&23.584/0.799&23.885/0.797&\textbf{26.550}/\textbf{0.883}\\
		\hline
		\hline
		Images:  &\multicolumn{7}{c|}{${\rm{SR}}=50\%$}\\
		\hline
		baboon&24.263/0.858 &21.837/0.764&23.547/0.839&23.004/0.812&23.704/0.839&23.099/0.816&\textbf{24.588}/\textbf{0.865}\\
		monarch&28.834/0.976 &23.558/0.927&26.256/0.957&24.089/0.932&24.079/0.935&24.587/0.938&\textbf{29.015}/\textbf{0.977}\\
		airplane&29.525/0.928 &26.626/0.835&28.217/0.922&26.768/0.822&27.195/0.869&27.438/0.850&\textbf{29.705}/\textbf{0.931}\\
		peppers&31.622/0.987&27.287/0.966&29.461/0.980&27.936/0.971&28.171/0.973&28.749/0.976&\textbf{31.675}/\textbf{0.989}\\  
		sailboat&27.541/0.950&24.723/0.890&26.251/0.932&25.000/0.892&25.525/0.910&25.414/0.902&\textbf{27.816}/\textbf{0.953}\\
		lena&30.749/0.980&27.385/0.962&29.360/0.975&27.476/0.962&28.189/0.968&28.128/0.967&\textbf{30.991}/\textbf{0.982}\\
		panda&30.414/0.914 &27.746/0.856&29.458/0.905&27.985/0.853&28.438/0.880&28.438/0.865&\textbf{30.548}/\textbf{0.916}\\
		barbara&28.259/0.924 &26.838/0.884&27.836/0.918&27.252/0.888&27.891/0.905&27.719/0.897&\textbf{29.041}/\textbf{0.931}\\
		\hline  
	\end{tabular}}   
	\label{tabE_1}  
\end{table}
\begin{figure*}[htbp]
	\centering
	\subfigure[baboon]{\includegraphics[width=16.5cm,height=10cm]{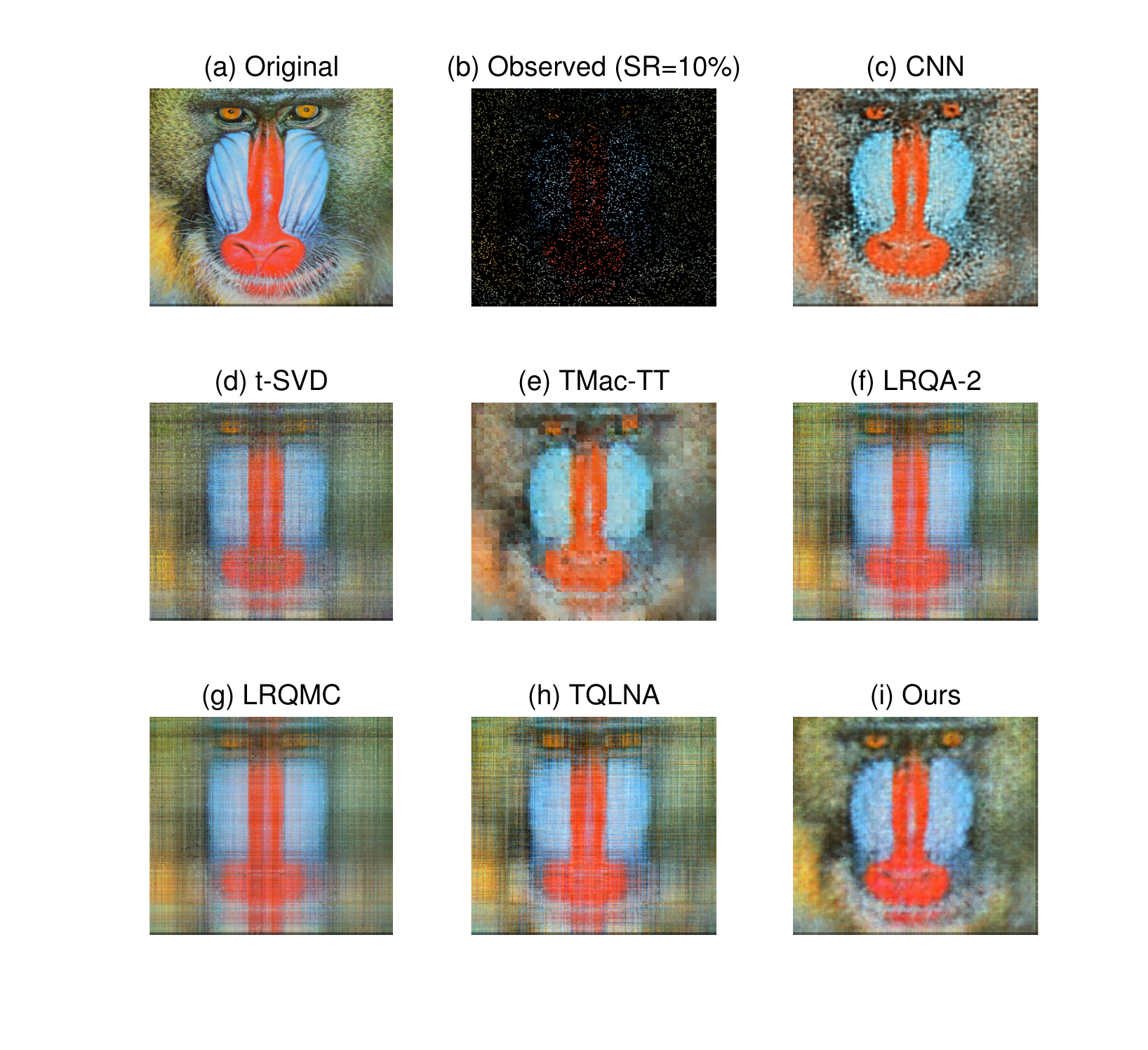}}
	\subfigure[monarch]{\includegraphics[width=16.5cm,height=10cm]{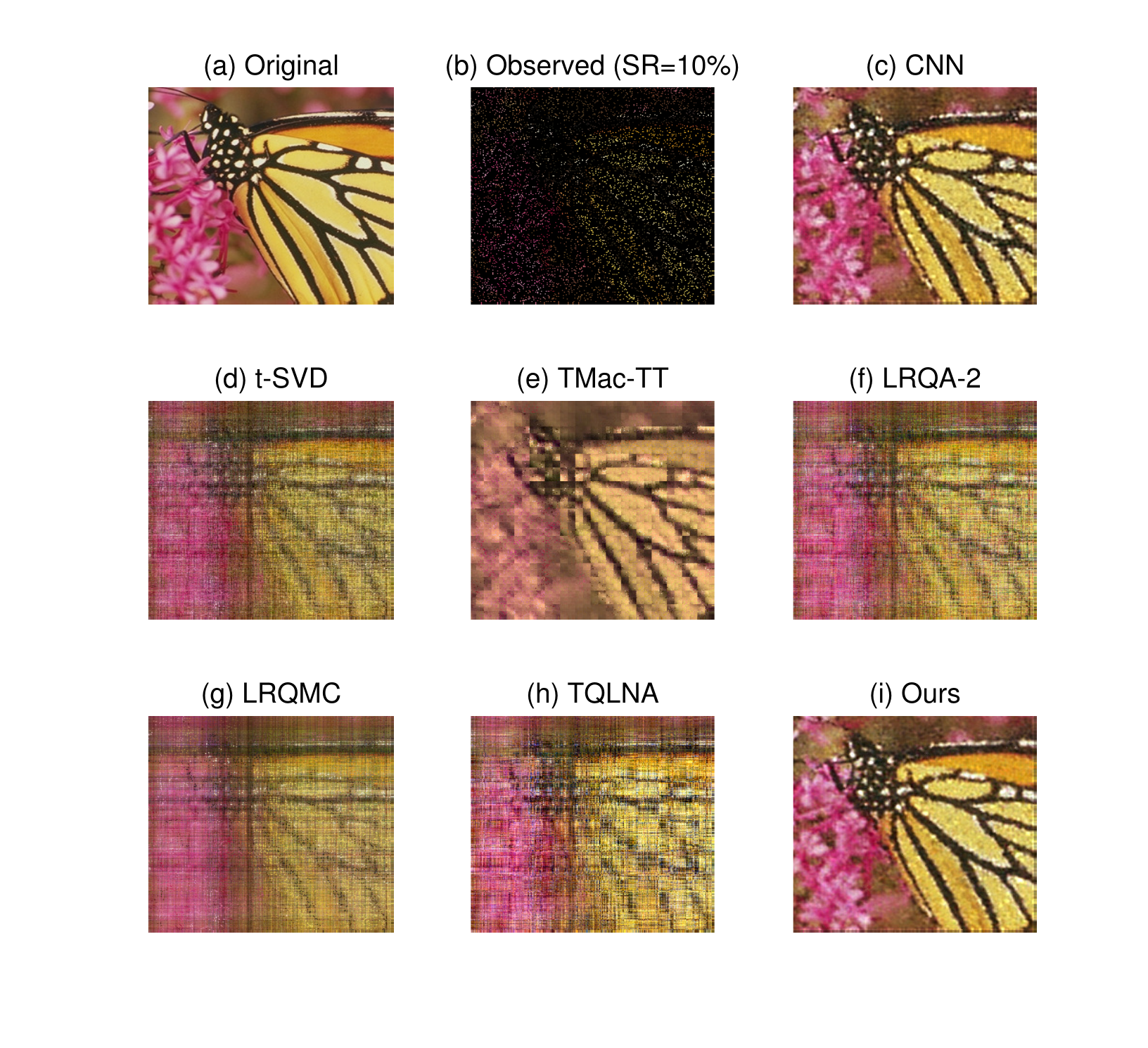}}
	\caption{Recovered two color images (baboon and monarch) for random missing with ${\rm{SR}} = 10\%$.}
	\label{im_Es1}
\end{figure*}
\begin{figure*}[htbp]
	\centering
	\subfigure[airplane]{\includegraphics[width=16.5cm,height=10cm]{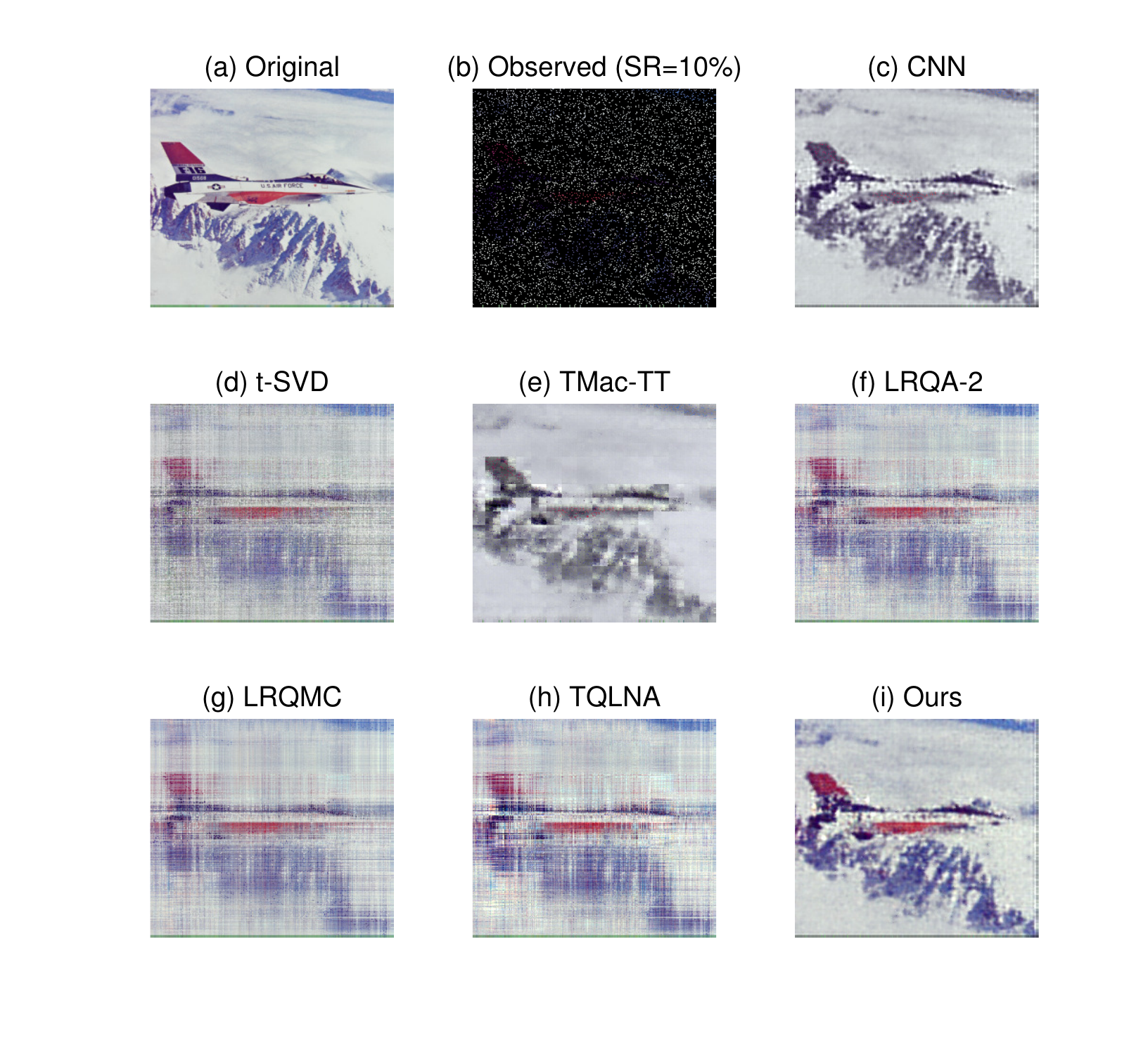}}	
\subfigure[panda]{\includegraphics[width=16.5cm,height=10cm]{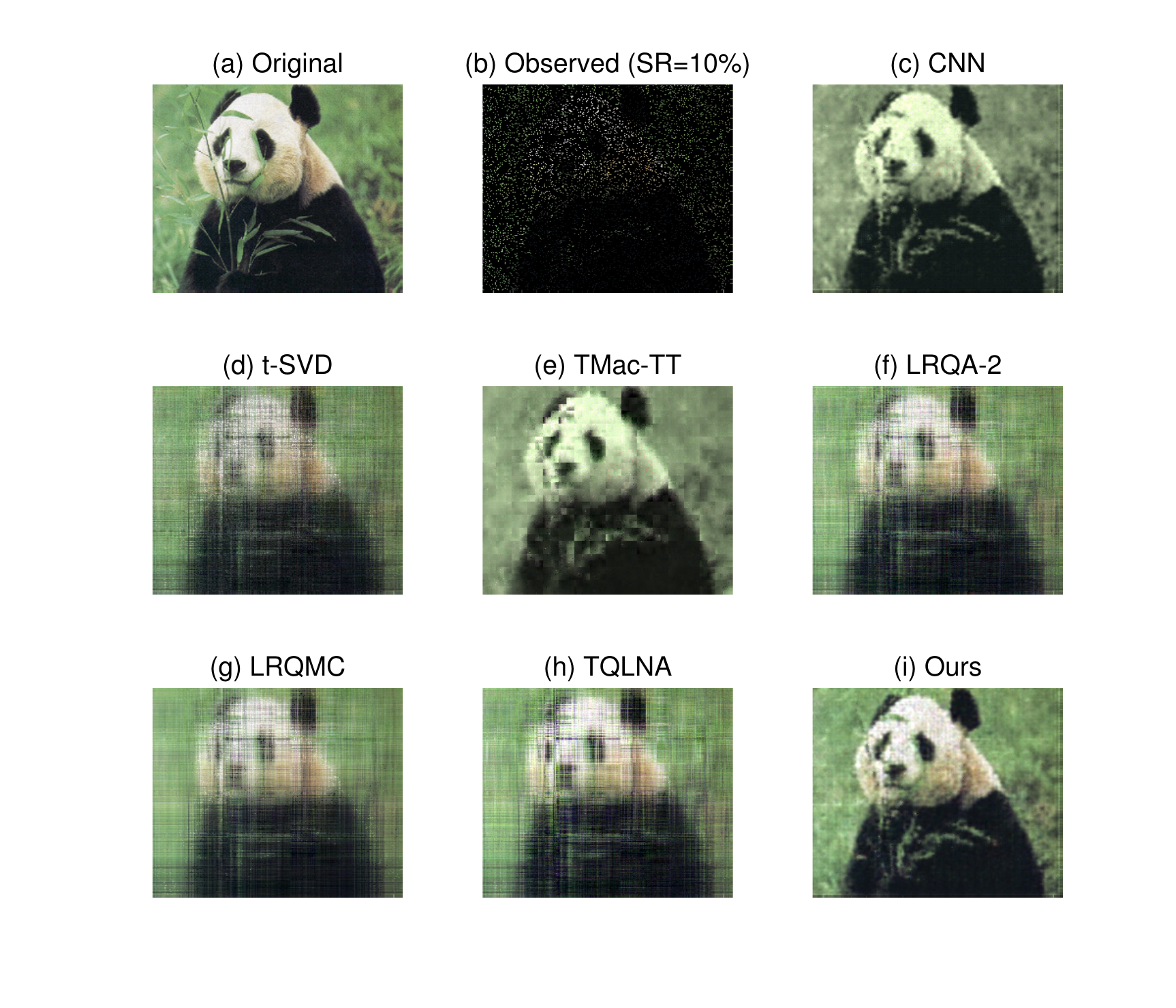}}
	\caption{Recovered two color images (airplane and panda) for random missing with ${\rm{SR}} = 10\%$.}
	\label{im_Es2}
\end{figure*}
\begin{figure*}[htbp]
	\centering
	\subfigure[airplane]{\includegraphics[width=16.5cm,height=10cm]{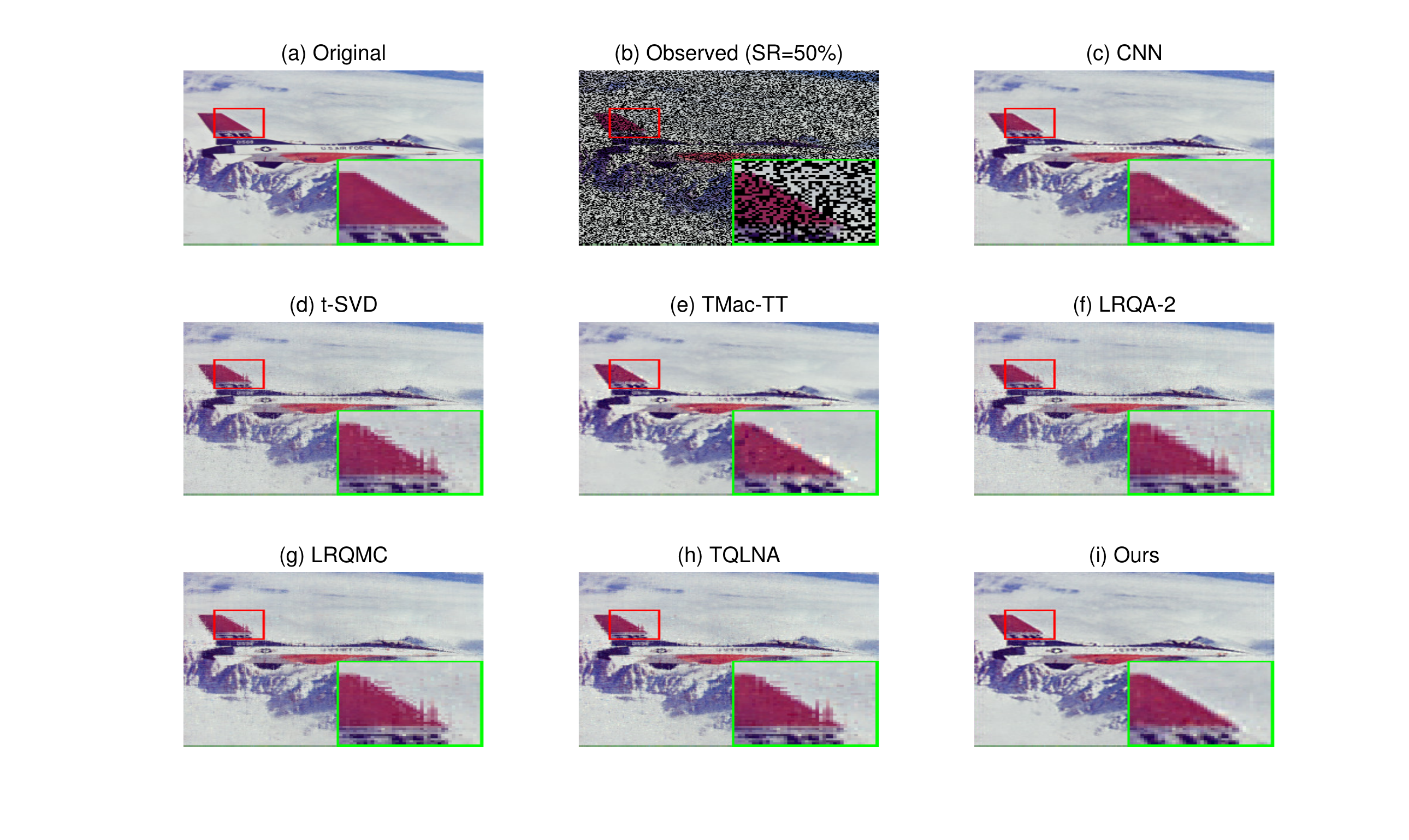}}
	\subfigure[peppers]{\includegraphics[width=16.5cm,height=10cm]{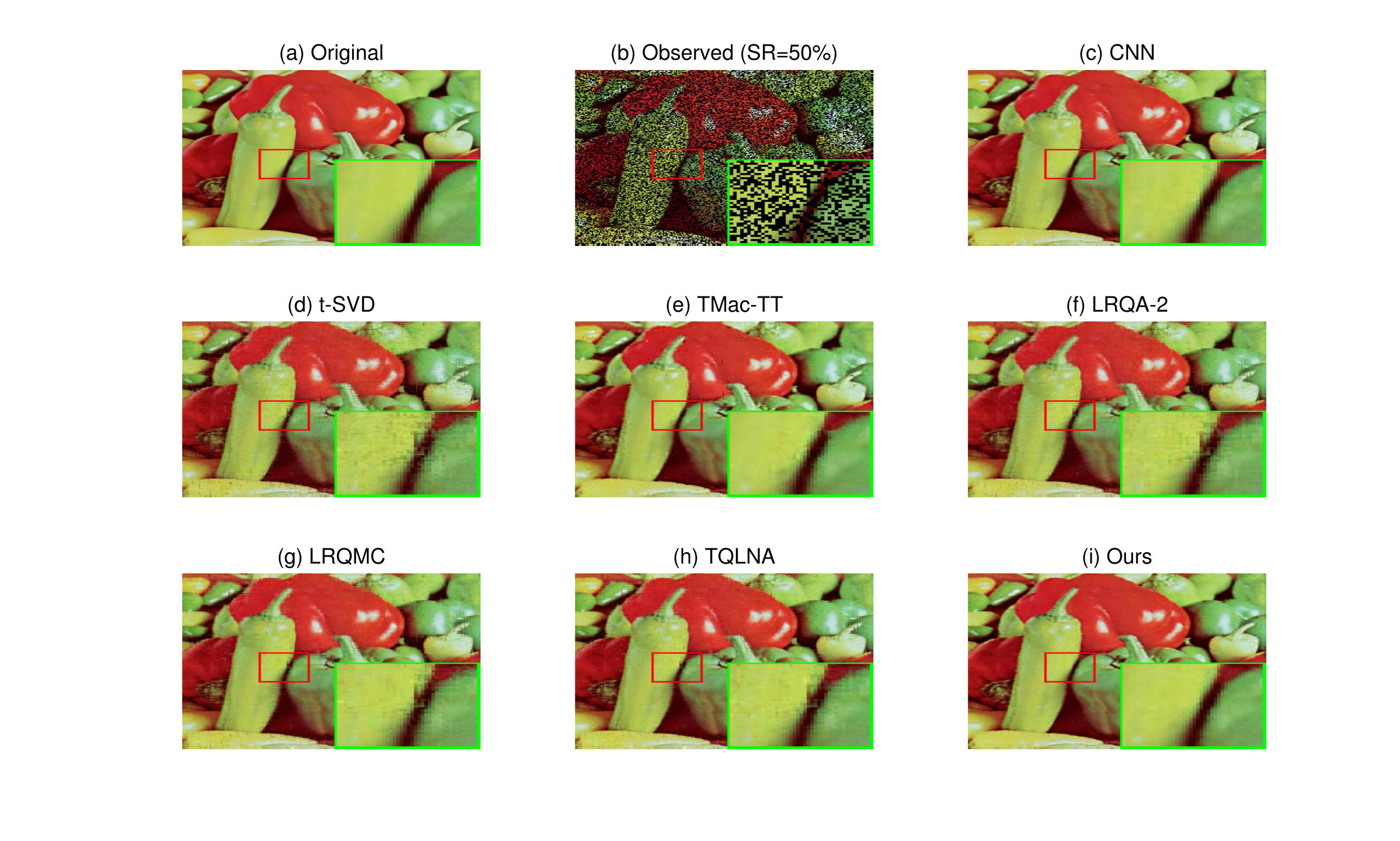}}	
	\caption{Recovered two color images (airplane and peppers) for random missing with ${\rm{SR}} = 50\%$.}
	\label{im_Es3}
\end{figure*}
\begin{figure*}[htbp]
	\centering
	\subfigure[lena]{\includegraphics[width=16.5cm,height=10cm]{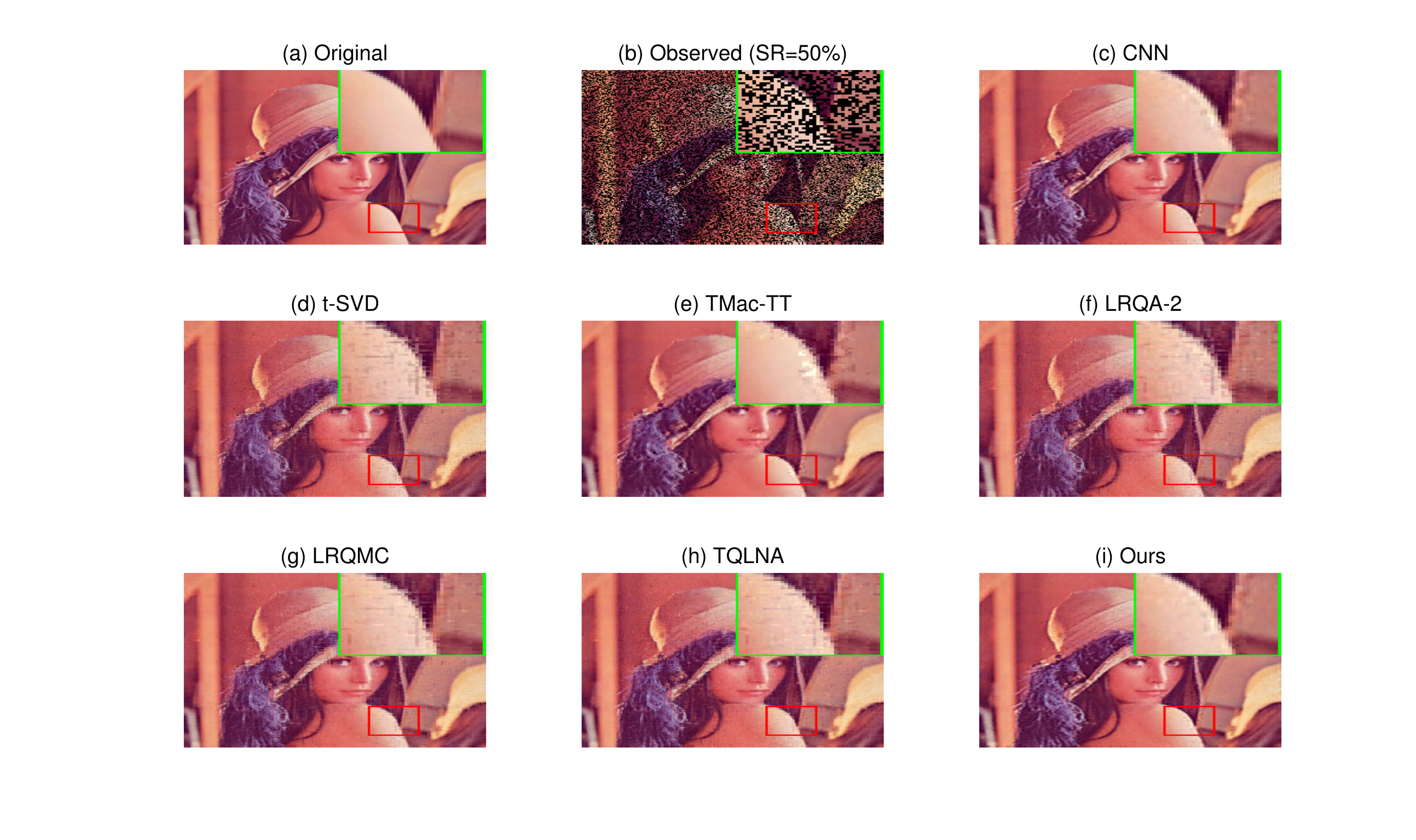}}
	\subfigure[barbara]{\includegraphics[width=16.5cm,height=10cm]{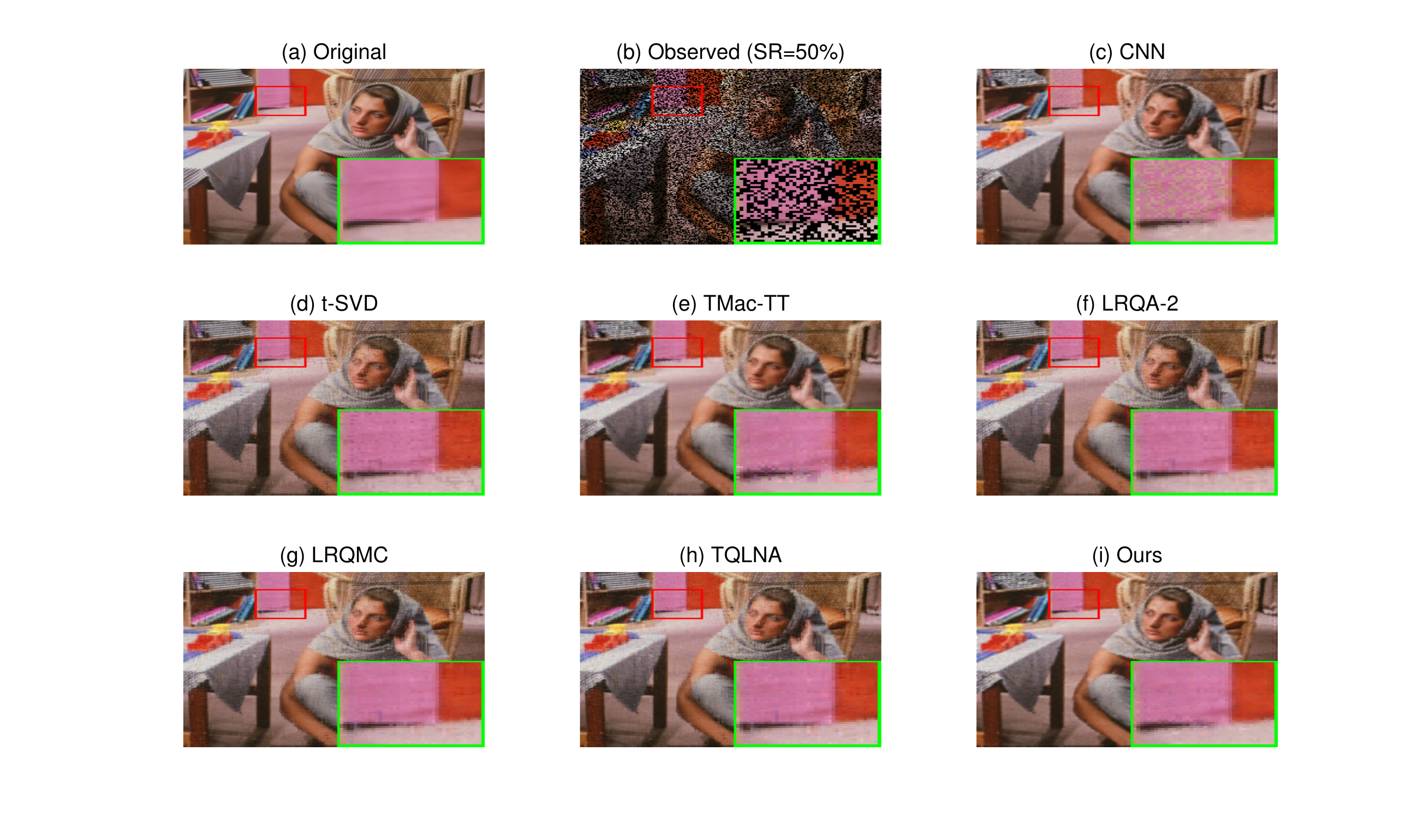}}	
	\caption{Recovered two color images (lena and barbara) for random missing with ${\rm{SR}} = 50\%$.}
	\label{im_Es4}
\end{figure*}
\begin{figure*}[htbp]
	\centering
	\subfigure[baboon]{\includegraphics[width=16.5cm,height=10cm]{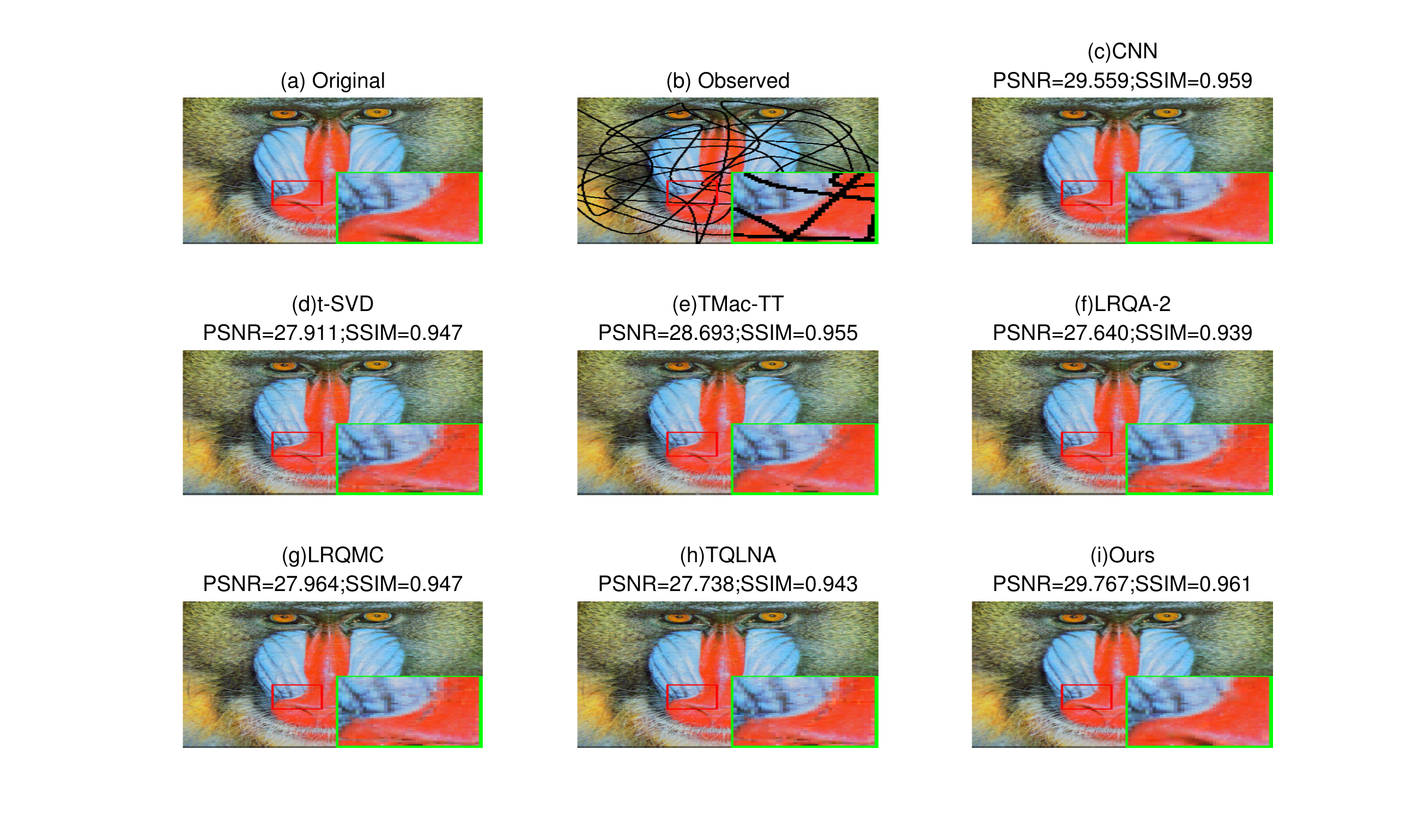}}
	\subfigure[lena]{\includegraphics[width=16.5cm,height=10cm]{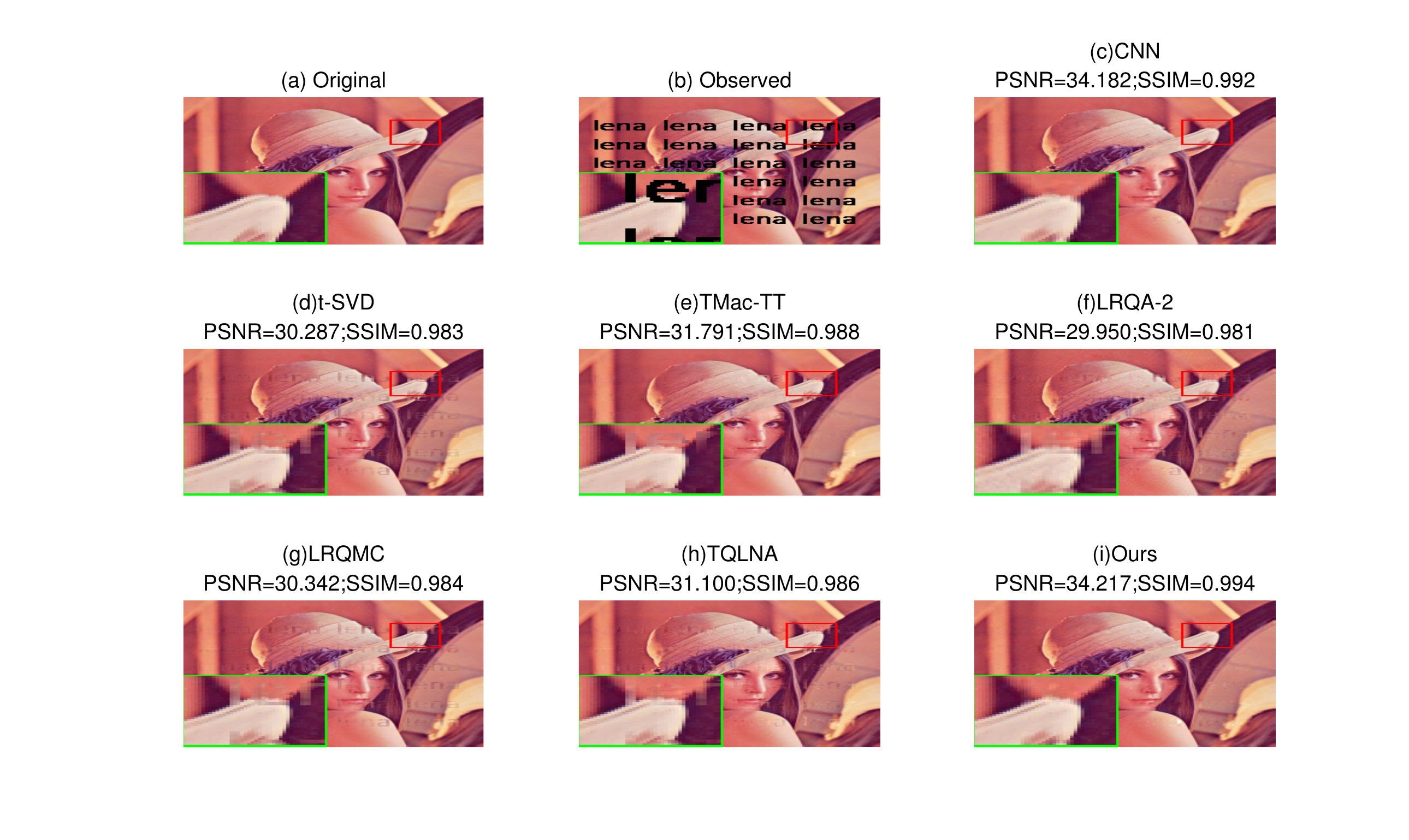}}	
	\caption{Recovered two color images (baboon and lena) for structural missing pixels.}
	\label{im_Es5}
\end{figure*}

\subsection{Results Analysis}
Table \ref{tabE_1} lists the PSNR and SSIM values of different methods on the eight color images with three levels of sampling rates. Figure \ref{im_Es1} and Figure \ref{im_Es2} display the recovered results of four color images by different methods for random missing with ${\rm{SR}} = 10\%$. Figure \ref{im_Es3} and Figure \ref{im_Es4} display the recovered results of four color images by different methods for random missing with ${\rm{SR}} = 50\%$. Experimental results of two kinds of structural missing cases are given in Figure \ref{im_Es5}. From all the experimental results, we can observe
and summarize the following points: 
\begin{itemize}
	\item Our QCNN-based color image inpainting method has advantages over the CNN-based method, especially in the case of a large number of lost pixels (\emph{e.g.}, SR=10\%), the advantages of our QCNN-based method are obvious. As shown in Figures \ref{im_Es1} and \ref{im_Es2}, when SR=10\%, the color images generated by the CNN-based method shows an obvious color difference compared with the original images. In addition, our QCNN-based method has advantages over CNN-based methods in preserving color image details (\emph{see} Figures \ref{im_Es3} and \ref{im_Es4}). The QCNN-based method has advantages over the CNN-based method mainly because the CNN-based method, for each kernel, simply merges the RGB three channels to sum the convolution results up, without considering the complicated interrelationship between different channels. This may result in the loss of important structural information.		
	\item Compared with existing methods based on low-rank approximation of quaternion matrices, our QCNN-based method replaced the explicit regularization term with an implicit prior learned by the QCNN. Therefore, our QCNN-based method has obvious advantages over the existing methods based on quaternion matrix low-rank approximation, both in terms of evaluation indicators and visually (\emph{see} Table \ref{tabE_1} and Figures \ref{im_Es1}-\ref{im_Es5}).
\end{itemize}

\section{Conclusion}
\label{sec:5}
For color image inpainting, this paper has proposed a quaternion matrix completion method using untrained QCNN. This approach enhances the quaternion matrix completion model by substituting the explicit regularization term with an implicit prior, which is acquired through learning by the QCNN. This method eliminates the need for researchers to spend time searching for appropriate regularization terms and designing intricate optimization algorithms for quaternion matrix completion models. From the experimental results, it can be seen that the method is very competitive with the existing quaternion matrix completion methods in the task of color image inpainting.

This paper represents the first attempt to apply QCNN to the quaternion matrix completion problem, and may therefore provide new insights for researchers exploring quaternion matrix completion methods, as well as other quaternion matrix optimization problems.

\section*{Acknowledgment}
This work was supported by University of Macau (File no. MYRG2019-00039-FST, MYRG2022-00108-FST ), Science
and Technology Development Fund, Macau S.A.R (File no.FDCT/0036/2021/AGJ).

\bibliographystyle{unsrt}
\bibliography{Myreference}
\end{document}